\documentclass[preprint,11pt,authoryear]{elsarticle}
\usepackage{amsmath,amssymb,amsthm}
\usepackage{graphicx}
\usepackage{xcolor}

\newtheorem{theorem}{Theorem}[section]

\theoremstyle{definition}

\theoremstyle{remark}
\newtheorem{remark}[theorem]{Remark}

\numberwithin{equation}{section}

\begin{document}

\title{A New Control Law for TS Fuzzy Models: Less Conservative LMI Conditions by Using Membership Functions Derivative\tnoteref{t1}}
\author[1]{Leonardo Amaral Mozelli\corref{cor1}}
\ead{lamoz@ufmg.br}
\author[1]{Victor Costa da Silva Campos}
\ead{kozttah@ufmg.br}
\cortext[cor1]{Corresponding author}
\affiliation[1]{
organization={Department of Electronics Engineering, Universidade Federal de Minas Gerais},
addressline={6627 Ave. Pres. Antônio Carlos, Pampulha},
postcode={31270-901},
city={Belo Horizonte},
country={Brazil}}
\tnotetext[t1]{This work has been supported by Brazilian agencies CAPES and CNPq.}








\begin{abstract}
This note proposes a new type of Parallel Distributed Controller (PDC) for Takagi-Sugeno (TS) fuzzy models. Our idea consists of using two control terms based on state feedback, one composed of a convex combination of linear gains weighted by the normalized membership grade, as in traditional PDC, and the other composed of linear gains weighted by the time-derivatives of the membership functions. We present the design conditions as Linear Matrix Inequalities, solvable through numerical optimization tools. Numerical examples are given to illustrate the advantages of the proposed approach, which contains the the traditional PDC as a special case.
\end{abstract}

\maketitle

\section{Introduction}

The control theory community witnessed a widespread application of Takagi-Sugeno (TS) fuzzy models since its inception, almost 40 years ago \citep{Takagi:ITSC:85}. In the early development stages, controllers were designed in an ad-hoc manner, with stability certificates being provided after the design \citep{Tanaka:FSS:92}. With the advent of efficient numerical techniques  and the realization that stability and stabilization conditions have characteristics of convex optimization, there was a paradigm shift in the field from the 90's and onwards \citep{Nguyen:ICIM:2019}. The field progressively gravitated towards the investigation of conditions based on Lyapunov functions that could be recasted as Linear Matrix Inequalities (LMIs). 

Over the last two decades, many advances have been promoted by using special functions, besides the quadratic ones: piecewise/switching \citep{CherifiAMC2019}; fuzzy \citep{YehAMC2008,NguyenAMC2022}; multiple \citep{LuoAMC2014}; polynomial; integral-type \citep{Mozelli:IS:09}, and so on.  The reader is refereed to the recent survey on this topic by \citet{Lam:EAAI:2019} and references therein. 

We focus the analysis on the so called Fuzzy Lyapunov Function (FLF). The idea is to use the information of the normalized membership grade to express the overall functions as a combination of local quadratic Lyapunov functions. 

The FLF is advantageous because it considers the rate of change among the local models that composed the global TS model. The way the quadratic function operates, the stability certificate must be strong enough for any rate of change.  From the numerical point of view, considering a TS model with the same number of rules, there are more decision variables available in the optimization problem with the FLF than compared to case of the quadratic stability, making it less complicated to find a solution. 

A disadvantage of the FLF is the necessity to consider the time-derivative of the membership functions in tractable manner in the LMI framework. Many alternatives have been proposed \citep{Jadbabaie:IFAC:99,Mozelli:A:09,Mozelli:IS:09,Bernal:ITFS:2010,Faria:IJSS:2013,Lam:BOOK:2016,Mozelli:IJRNC:19,Vieira:AMC:23}, to the point that this factor is not impending from the conservatism and numerical complexity points of view. Another disadvantage is the fact that the stability conditions are locally valid. However, since most TS fuzzy models have limited universe of discourse, this factor is not critical if large domains of attraction can be established \citep{Pan:TFS:2012,Lee:Inf:2012,Lee:Cyb:2013,Marquez:FSS:2017,Campos:IFAC:2017}. 

In this paper, by harnessing the advancements of the last decades, in terms of numerically expressing stability conditions based on FLF in the LMI framework, we propose a new controller structure that includes the traditional Parallel Distributed Compensator (PDC) as a special case. This new controller is less conservative than previous approaches and can guarantee larger domains of attraction. The rationale motivating this new controller is that the information regarding rate of change of the membership grades, when explicitly considered in the control law, can promote an anticipatory behavior, dampening the transient response. 

The remainder of this paper is organized as follows: Section~\ref{sec:prelim} presents mathematical notation, background, and techniques necessary in this paper; Section~\ref{sec:main} concentrates the main results of this paper, including the Theorems for stabilization, LMIs for designing the new type of controller and numerical examples to illustrate their effectiveness; finally conclusions and futures avenues of research are paved in Section~\ref{sec:conclu}.

\textbf{Notation:} Let $M \in \mathbb{R}^{r \times p}$ be a real matrix composed of $p$ column vectors $m^j \in \mathbb{R}^r$, which, in turn, are constituted by $r$ elements $m^j_k$. For symmetric matrices, $M<0,(>0)$ is definite negative (positive), and $M'$ is the transpose. The convex hull of a set $H = \{v^1,v^2,\ldots,v^p\},v^j\in \mathbb{R}^r$ is denoted by $\text{co}(H)$. The symbol $\bullet$ indicates transposed terms in block matrices. The dependence of some variables on the continuous time $t\in\mathbb{R}$ is omitted for the sake of presentation, e.g., $x(t), u(t), h(t)$. 

\section{Preliminaries}\label{sec:prelim}

Consider Takagi-Sugeno (TS) fuzzy models:

\begin{equation}\label{eq:TS-model}
    \dot{x} = A(h)x +B(h)u,
\end{equation}

\noindent where $x$ is the state vector, $u$ is the control signal, $h = [h_1~h_2~\cdots~h_r]'$ is the vector composed of membership functions, and $A(h),B(h)$ comes from the fuzzy combination of $r$ local models, abiding by the convex property:

\begin{equation}
    A(h) = \sum^r_{i=1}h_i A_i,~h_i\geq 0,~\sum^r_{i=1}h_i=1.
\end{equation}

The PDC controller is given by similar combination:

\begin{equation}\label{eq:tradPDC}
    u = K(h)x = \sum^r_{i=1}h_iK_ix,
\end{equation}

\noindent leading to the closed-loop form:

\begin{equation}\label{eq:closed_loop_old}
    \dot{x} = [A(h)+B(h)K(h)]x = \sum^r_{i=1}\sum^r_{j=1}h_ih_j [A_i+B_iK_j]x.
\end{equation}

Over the last two decades, the usage of the information regarding the rate of variation among local models has proven fruitful for stability analysis and stabilization, in the Lyapunov sense, for this kind of nonlinear dynamics. Those results are concerned with finding a scalar function:

\begin{equation}\label{eq:lyap_fuzzy}
    V(x,h) = x'P(h)x >0,~\forall x \neq 0, 
\end{equation}

\noindent whose rate of change along the solutions 

\begin{equation}\label{eq:der_lyap_fuzzy}
    \dot{V}(x,h) = \dot{x}'P(h)x + x'P(h)\dot{x}+ x'P(\dot{h})x <0, 
\end{equation}

\noindent where $P(\dot{h})$ explicitly depends on the time derivative of the membership functions, according to:

\begin{equation}\label{eq:prop_der_mf}
    P(\dot{h}) = \sum^r_{i=1}\dot{h}_i P_i,~\underline{\phi}_i \leq \dot{h}_i \leq \bar{\phi}_i,~\sum^r_{i=1}\dot{h}_i=0.
\end{equation}

\noindent with lower and upper bounds $\underline{\phi}_i, \bar{\phi}_i$.

Replacing \eqref{eq:closed_loop_old} into \eqref{eq:der_lyap_fuzzy} results in:

\begin{equation}\label{eq:der_lyap_fuzzy_cl}
    \dot{V}(x,h) = x'\{ [A'(h)+K'(h)B'(h)]P(h) + P(h)[A(h)+B(h)K(h)] + P(\dot{h})\} x
\end{equation}

From conditions \eqref{eq:lyap_fuzzy} and \eqref{eq:der_lyap_fuzzy_cl} it is possible to find the stabilizing gains of the PDC controller and the matrices of the function that provides a certificate of stability in the Lyapunov sense. The idea is to recast these conditions as, numerically solvable, LMI constraints. However there are three main issues in developing LMI constraints: the multiplication of decision variables $P(h)K(h)$; the inclusion of the term $P(\dot{h})$; and imposing the constraints on the time derivative bounds.

To solve the first issue, some works proposed the use of slack matrices variables to decouple matrices $P(h)$ from $K(h)$ \citep{Guerra:A:04,Tanaka:ITFS:07,Mozelli:IS:09}. \citet{Mozelli:CTA:10} demonstrate that towards this objective, Finsler's Lemma, the null term summation and the descriptor system approaches are equivalently effective. 

To solve the description of the term $P(\dot{h})$ there are many approaches, with distinct degrees of conservatism and numerical complexity. \citet{Jadbabaie:IFAC:99} sought a solution by combining the extreme values of \eqref{eq:prop_der_mf}, which is conservative and leads to $2^r$ constraints. \citet{Mozelli:A:09} observe the more conservative scenario and took advantage of the property of zero sum of the time-derivatives, last term in \eqref{eq:prop_der_mf}, to incorporate slack variables and reduce conservatism whilst keeping the number of constrains amendable. However, the exact description is proposed by  \cite{Geromel:SCL:06} and \cite{Chesi:A:07}, independently, which consists in realizing that the time-derivatives of the membership functions lie in a manifold of dimension $r-1$ given by the intersection of a hiperrectangle:

\begin{equation}\label{eq:rectangle}
    \dot{h} \subset \mathcal{B} \equiv \{\dot{h} \in \mathbb{R}^r: \underline{\phi}_i \leq \dot{h}_i \leq \bar{\phi}_i,~\forall i=1,2,\ldots,r \}
\end{equation}

\noindent defined by the bounds of the derivatives, and a hiperplane defined by the zero sum property:

\begin{equation}\label{eq:plane}
    \dot{h} \subset \mathcal{\pi} \equiv \{\dot{h} \in \mathbb{R}^r: e'\dot{h}=0 \}
\end{equation}

\noindent where $e' = [1~1~\cdots~1] \in \mathbb{R}^r$.

Therefore, in this paper we adopt the procedure of considering 

\begin{equation}\label{eq:exact_poly}
    \dot{h} \in \mathcal{D}_p = \mathcal{B} \cap \mathcal{\pi} = \text{co}(v^1,~v^2,~\ldots,v^p) \equiv \{v^j\in\mathbb{R}^r:\underline{\phi}_i \leq v^j_k\leq \bar{\phi}_i,~ e'v^j=0\}.
\end{equation}

Finally, in order to derive suitable LMI conditions that guarantee that a desired bound for the time derivative will be respected in closed loop, \cite{Pan:TFS:2012} proposed to employ Young's inequality, as well as bounds for the state and control action, whereas \cite{Lee:Inf:2012} proposed to directly bound the time derivative using a level set of the Lyapunov function. Barring the existence of extra structure on the membership functions, \cite{Campos:IFAC:2017} showed that the approach used in \cite{Lee:Inf:2012} usually yields better results, and as such will be a starting point for the conditions proposed in this paper.

Based on this body of literature, many elegant numerical solutions become available to solve the design of PDC controllers based on fuzzy Lyapunov functions in the last two decades. In the sequel, supported by these advancements we introduce a new kind of PDC controller. 

\section{Main Results}\label{sec:main}

This section presents a new PDC controller that includes a control action depending on the time derivative of the membership function. New LMI conditions, that include previous PDC controllers as special cases, are offered based on numerical solutions to consider the time-derivative and avoid dependencies on the decision variables. 

\subsection{Base Conditions}

The new PDC controller is composed of two actions:

\begin{equation}\label{eq:newPDC}
    u = K(h)x + L(\dot{h})x = \left[K(h) + L(\dot{h})\right]x.
\end{equation}

One action is the same as in  the traditional PDC controller and the new action depends on the time derivative of the membership functions. Notice that the traditional PDC controller can be recovered by setting $L(\dot{h})$ to zero. 

Using this  controller, a new closed-loop is given by:

\begin{equation}\label{eq:closed_loop_new}
    \dot{x} = [A(h)+B(h)K(h)+B(h)L(\dot{h})]x 
\end{equation}

Replacing \eqref{eq:closed_loop_new} into \eqref{eq:der_lyap_fuzzy} results in:

\begin{align}\label{eq:der_lyap_fuzzy_new_cl}
    \dot{V}(x,h) = &~ x'[A'(h)+K'(h)B'(h)+L'(\dot{h})B'(h)]P(h)x  \\
                & +x'P(h) [A(h)+B(h)K(h)+B(h)L(\dot{h})]x + x'P(\dot{h})x <0    
\end{align}

From this new condition, some numerical solutions can be employed towards the development of a set of LMI conditions.

First the multiplication of decision variables will be addressed.

\begin{theorem}
    Let $\alpha>0$ be a given scalar. The TS fuzzy model \eqref{eq:closed_loop_new} is asymptotically stable if the following conditions are satisfied:

    \begin{equation}\label{eq:theo:linear}
        T(h)> 0,~ M(h,\dot{h})<0
    \end{equation}

\noindent where 

\begin{align*} 
    &M{(h,\dot{h})} =  \\ 
    &\begin{bmatrix}
        \begin{matrix}
        T(\dot{h}) +A(h)R+B(h)S(h)+B(h)U(\dot{h}) +\\R'A'(h)+S'(h)B'(h)+U'(\dot{h})B'(h)
        \end{matrix}
        & \bullet \\
        &\\
        T(h)-R' +\alpha A(h)R +\alpha B(h)S(h) +\alpha B(h)U(\dot{h}) & -\alpha(R+R')
    \end{bmatrix}    
\end{align*} 
\end{theorem}

\begin{proof}
    Rewrite \eqref{eq:der_lyap_fuzzy} as: 
    \begin{equation}
        \dot{V}(x,h) = 
        \xi' 
        \begin{bmatrix}
            P(\dot{h}) & \bullet \\
            P(h) & 0
        \end{bmatrix}
        \xi <0
    \end{equation}
\noindent where $\xi' = [x'~\dot{x}]$. Observing the following identity:

\begin{equation}
    2[x'M_1+\dot{x}'M_2] \times \{ -\dot{x} +[A(h)+B(h)K(h)+B(h)L(\dot{h})]x\} = 0
\end{equation}
\noindent it follows that    $\dot{V}(x,h) = \xi '\Xi \xi < 0$ where:

\begin{equation}
        \Xi = 
        \begin{bmatrix}
        \begin{matrix}
            P(\dot{h}) +M_1[A(h)+B(h)K(h)+B(h)L(\dot{h})] \\+[A'(h)+K'(h)B'(h)+L'(\dot{h})B'(h)]M'_1 
            \end{matrix}
            & \bullet \\
            &\\
            P(h) -M'_1 +M_2[A(h)+B(h)K(h)]& -(M_2 + M_2')
        \end{bmatrix}
\end{equation}

Making the particular choice $M_1' = R^{-1},~ M_2 = \alpha M_1$ and using the similarity transformation $\Theta = \text{diag}(M^{-1}_1,M^{-1}_1)$, linear conditions are obtained by setting
$M(h,\dot{h}) = \Theta \Xi \Theta '$  where
\begin{equation}
    T(h) = R'P(h)R,~ S(h)=K(h)R,~ U(h) = L(h)R.
\end{equation}

Therefore, it suffices to impose \eqref{eq:theo:linear} to guarantee that $\dot{V}(h,\dot{h})<0$ and $P(h)>0$, which concludes the proof. 
\end{proof}

This Theorem is not yet an LMI, since there is the dependency on the information regarding membership functions. In the sequel, the properties of the membership functions  and its time derivatives will be used to arrive at sufficient and finite conditions, which are solvable numerically. 

\begin{theorem}
    Let $\alpha>0$ be a scalar given. The TS fuzzy model \eqref{eq:closed_loop_new} is asymptotically stable if the following conditions are satisfied:

    \begin{equation}\label{eq:theo_suff}
        T_i> 0,~ M_{i,i,\ell}<0, \forall i \in \mathcal{R}, \ell\in\mathcal{V},~ M_{i,j,\ell}+M_{j,i,\ell}<0,~\forall i,j \in \mathcal{R}, \ell\in\mathcal{V}, i<j
    \end{equation}

\noindent where 

\begin{align*} 
    &M_{i,j,\ell} =  \\ 
    &\begin{bmatrix}
        \sum_{k=1}^r v^\ell_k( T_k +B_iU_k +U'_kB'_i) +A_iR+B_iS_j+R'A'_i+S'_jB'_i& \bullet \\
        T_i-R' +\alpha A_iR +\alpha B_iS_j +\alpha\sum_{k=1}^r v^\ell_k B_iU_k& -\alpha(R+R')
    \end{bmatrix}    
\end{align*} 

\noindent and $v^\ell_k$ are the elements from the following matrix:

\begin{equation}\label{eq:vertexH}
    H = 
    \begin{bmatrix}
    v^1,  & v^2, & \cdots, & 
    \begin{pmatrix}
    v^\ell_1\\
    v^\ell_2 \\
    \vdots \\
    v^\ell_r
    \end{pmatrix},
    \cdots,
    v^p
    \end{bmatrix}
\end{equation}

\noindent representing the convex hull of \eqref{eq:exact_poly}.

\end{theorem}

\begin{proof}
This proof is based on Theorem~1. Notice that \eqref{eq:theo:linear} can be rewritten as:

\begin{align}
    T(h) &= \sum^r_{i=1}h_iT_i > 0 \label{eq: partial1}\\
    M(h,\dot{h}) &= \sum^r_{i=1}\sum^r_{j=1}h_ih_jM_{i,j}(\dot{h}) <0 \\
                 &=  \sum^r_{i=1}h_i^2M_{i,i}(\dot{h}) + \sum^r_{i<j}h_ih_j [M_{i,j}(\dot{h})+M_{j,i}(\dot{h})] <0 \label{eq: partial2}
\end{align}

\noindent where

\begin{equation}
    M_{i,j}(\dot{h}) = 
    \begin{bmatrix}
        \begin{matrix}
        T(\dot{h}) +B_iU(\dot{h}) +U'(\dot{h})B'_i \\
        +A_iR+B_iS_j+R'A'_i+S'_jB'_i 
        \end{matrix}
        & \bullet \\
        &\\
        T_i-R' +\alpha A_iR +\alpha B_iS_j +\alpha B_iU(\dot{h})& -\alpha(R+R')
    \end{bmatrix} 
\end{equation}

To address the dependency on the rate of change $\dot{h}$, notice the property  \eqref{eq:prop_der_mf}. Given that the sum of the time-derivatives of the membership functions is always zero, they are confined to a $r-1$-dimensional manifold $\mathcal{D}_p$ \eqref{eq:exact_poly}. 

Since this is a convex region, there exists a matrix \eqref{eq:vertexH} that accumulates the vertices of its hull. Therefore, to impose conditions \eqref{eq: partial1} and \eqref{eq: partial2} it is sufficient to verify \eqref{eq:theo_suff} at the vertices, which concludes the proof.
\end{proof}

\begin{remark}
    Algorithms to construct matrix \eqref{eq:vertexH} are available in \citet{Mozelli:IJRNC:19} and \citet{Vieira:AMC:23}. As these works discuss, although this method (exact polytope) is the least conservative to include the information regarding the rate of change of membership functions, the numerical impact might be prohibitive for large number of rules, $r$. Trade-off solutions are presented, where there is a marginal increase in conservatism, with substantial reduction in the numerical burden. 
\end{remark}

\subsubsection{Numerical Example}

This section conducts some numerical comparisons to illustrate the advantages of the proposed controller. Consider a TS fuzzy model \eqref{eq:TS-model} with the parameters presented in \citet{Lazarini:I3A:2021}:

\[
    A_1 =
    \begin{bmatrix}
    3.6 & -1.6 \\ 6.2 & -4.3
    \end{bmatrix}, ~~
    A_2 =
    \begin{bmatrix}
    -a & -1.6 \\ 6.2 & -4.3
    \end{bmatrix},
    B_1 =
    \begin{bmatrix}
    -0.45 \\ -3
    \end{bmatrix}, ~~
    B_2 =
    \begin{bmatrix}
    -b \\ -3
    \end{bmatrix},
\]

The objective is to find stabilizing PDC controllers for the closed loop system considering the parametric space $a\in[0,10]$, $b\in[1,2]$ under distinct methodologies.

The investigations assumed $\alpha=0.04$, $\underline{\phi}_1 = \underline{\phi}_2 = -1$, and $\bar{\phi}_1 = \bar{\phi}_2 = 1$. The codes are implemented using the combination of MatLab, YALMIP \citep{Lofberg:CACSD:04}, and SeDuMi \citep{Sturm:OMS:99} as programming language, parser and solver, respectively. 

\begin{figure}[h]
\centering
\includegraphics[width=0.725\textwidth]{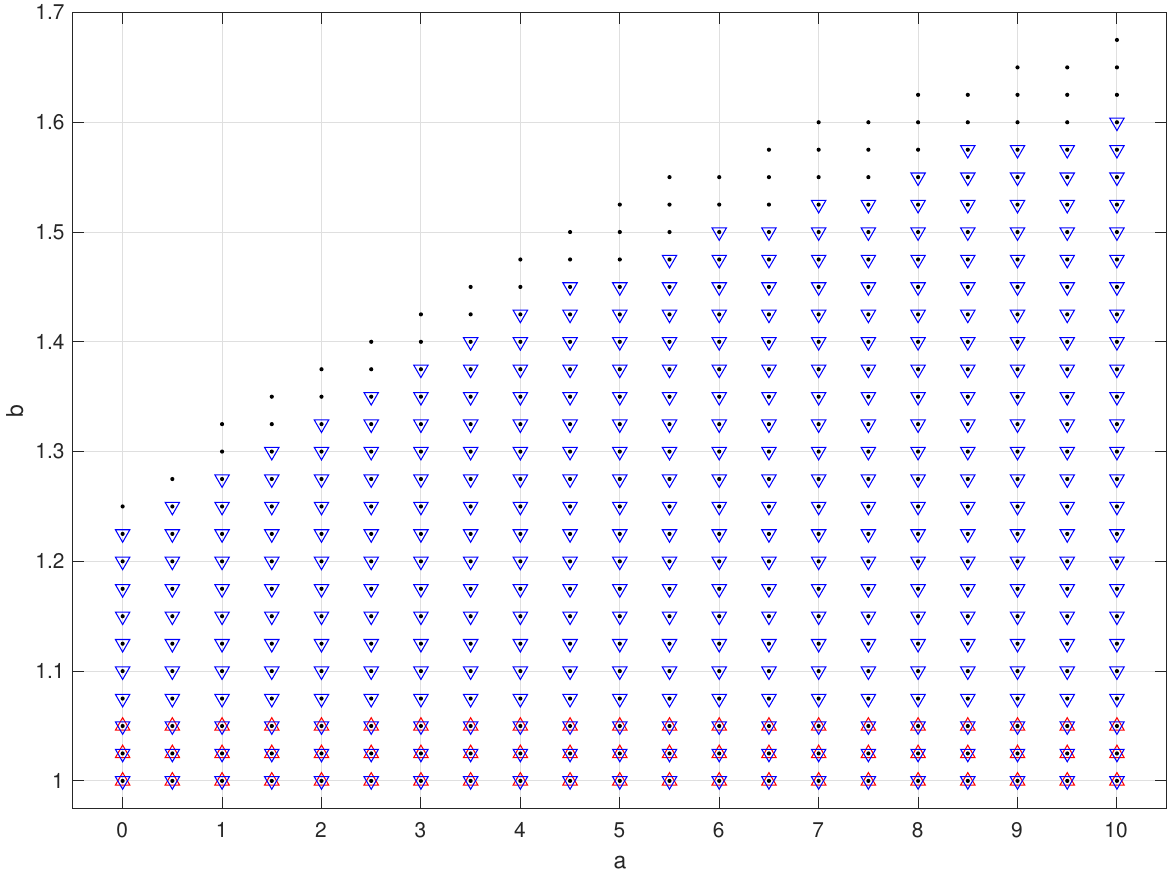}
\caption{Stabilization region according to different methodologies to design PDC controllers: traditional PDC controller with quadratic Lyapunov function (red triangle); PDC controller with fuzzy Lyapunov function, from \citet{Lazarini:I3A:2021} (blue triangle); proposed (black dot).}
\label{fig:stab_space}
\end{figure}

Figure~\ref{fig:stab_space} shows the comparison among three methodologies. Two are based on the traditional PDC controller \eqref{eq:tradPDC}, being that one based on quadratic Lyapunov function and the other based on the fuzzy Lyapunov function and PDC controller, from \citet{Lazarini:I3A:2021}. Finally, the remaining result is based on the new PDC controller \eqref{eq:newPDC}. Whenever the methodology is feasible for a given parametric pair, they are marked by red upward triangle, blue downward triangles and black dot, respectively.

Figure~\ref{fig:stab_space} confirms the superiority of considering a fuzzy Lyapunov function over the quadratic form. As the parameter $a$ increases, there is a clear trend of enlarging the stabilization for the parameter $b$, whereas for the quadratic Lyapunov function they remain the same. 

The proposed method, as expected, includes the result from the other PDC controller methodology.  Even though there are only two local rules, the proposed methodology is able to surpass a recent stability condition based on PDC controller for all parameter set investigated.

\subsection{Local Conditions}

In order to guarantee the constraints on the derivative bounds for the membership functions, local stabilization must be considered. Let us constrain the states to a hyper-rectangle:
\begin{equation} \label{eq:hyper_rect}
    \Upsilon = \{ x \in\mathbb{R}^n~ |~ |x_k|  \leq \bar{x}_k,~\forall k \in \{ 1,2,\cdots,n \} \} 
\end{equation}
\noindent where $x_k$ is the $k$-th coordinate of $x$.

A locally stable region is called the Domain of Attraction (DA), whose initial conditions converges to the origin without leaving the region:

\begin{equation}
    \Omega = \{ x(0) \in \mathbb{R}^n | \lim_{t \rightarrow \infty}x(t) \rightarrow 0\}
\end{equation}

An inner estimate of the DA can be found by computing a sublevel set of \eqref{eq:lyap_fuzzy}:

\begin{equation}
    \Upsilon \supset \Omega^*(c) = \{ x \in \mathbb{R}^n | x'P(h)x \leq c\}
\end{equation}

Since, in this work, we are only interested in the stabilization of the system, without loss of generality we will focus our attention to the 1-sublevel set of \eqref{eq:lyap_fuzzy}, $\Omega^*(1)$, and suitable conditions will ensure that it is a subset of the hyper-rectangle $\Upsilon$.

The time-derivative of the membership functions can be represented as:

\begin{equation}
    \dot{h}_\upsilon = \nabla h_\upsilon \dot{x} = \sum^s_{q=1} \tau^\upsilon_q \zeta^\upsilon_q \dot{x},~\forall \upsilon \in \mathcal{R}
\end{equation}

\noindent where $\zeta^\upsilon_q$ is a row vector and the nonlinear functions $\tau^\upsilon_q$ satisfy:

\begin{equation}
    \tau^\upsilon_q \in [0,1],~ \sum^s_{q=1} \tau^\upsilon_q =1.
\end{equation}

Bearing in mind these constraints, local stabilization conditions can be imposed by the following Theorem:

\begin{theorem}\label{theo:local_theo}
    Let $\alpha>0$ be a scalar given. The closed-loop TS fuzzy model \eqref{eq:closed_loop_new} is locally asymptotically stable, with $\Omega^*(1)$ an estimation of its domain of attraction, guaranteeing that inside of this region $|\dot{h}_\upsilon| \leq \phi_\upsilon$ and $\left|1 - \nabla h_\upsilon B(h)L_\upsilon x\right| \geq \mu_\upsilon$, if conditions \eqref{eq:theo_suff} are verified and: 
    \begin{align}
        &\begin{bmatrix} -T_i & R'e_k \\ e_k'R & -\bar{x}_k^2 \end{bmatrix} \leq 0, \; \forall i \in \mathcal{R}, \forall k \in \mathcal{N} \label{eq:omega1}\\
        &Q_{ii \upsilon q \ell} \leq 0, \; \forall i, \upsilon \in \mathcal{R}, \forall q \in \mathcal{S}_\upsilon, \forall \ell \in \mathcal{V} \label{eq:qii}\\
        &Q_{ij \upsilon q \ell} + Q_{ji \upsilon q \ell} \leq 0, \; \forall i, j> i, \upsilon \in \mathcal{R}, \forall q \in \mathcal{S}_\upsilon, \forall \ell \in \mathcal{V} \label{eq:qij}\\
        &X_{i\upsilon q} \leq 0 , \; \forall i, \upsilon \in \mathcal{R}, \forall q \in \mathcal{S}_\upsilon \label{eq:Xiuq}
    \end{align}
    with
    \begin{align}
        Q_{ij\upsilon q\ell} &= \begin{bmatrix}
            -T_i & \bullet \\
            \zeta^\upsilon_q \left(A_i R + B_i S_j + B_i \displaystyle \sum_{\substack{w=1\\ w\neq \upsilon}}^r v^\ell_w U_w\right) & -\mu_\upsilon^2 \phi_\upsilon^2
        \end{bmatrix}, \\
        X_{i\upsilon q} &= \begin{bmatrix}
            -T_i & \bullet \\
            \zeta^\upsilon_q B_i U_\upsilon & -(1 - \mu_\upsilon)^2
        \end{bmatrix}.
    \end{align}
    In addition to this, the estimated domain of attraction, $\Omega^*(1)$, is enlarged by maximizing $\log(\det(H))$, with $H$ a symmetric positive definite matrix, with the additional conditions
    \begin{align}
        -R-R'+T_i+H \leq 0, \; \forall i \in \mathcal{R}. \label{eq:lyapQF}
    \end{align}
    
\end{theorem}

\begin{proof}
    First, we impose the 1-sublevel set of the Lyapunov function, $\Omega^*(1)$, is contained in the local region, $\Upsilon$, by setting: 
    \begin{equation}\label{eq:initial_level1}
        P(h) > \dfrac{e_ke'_k}{\bar{x}_k^2},~\forall k\in\mathcal{N}.
    \end{equation}
\noindent where $e_k$ is a vector of zeros, except in the $k$-th coordinate, i.e.:
\begin{equation}\label{eq:vector_single_one}
    e_k = \begin{bmatrix}0&\cdots&0&1&0&\cdots&0\end{bmatrix}' \in \mathbb{R}^n
\end{equation}

Notice that multiplying \eqref{eq:initial_level1} by $x'$ and $x$ on the left and right, respectively, and assuming $x_k = \bar{x}_k$ results in:
\begin{equation}
    x'P(h)x > \dfrac{x_kx'_k}{\bar{x}_k^2} = 1
\end{equation}
which ensures that, on the border of the $\Upsilon$ region the Lyapunov function is equal or larger than one. Since the Lyapunov function is positive definite and continuous, this guarantees that $\Omega^*(1) \subset \Upsilon$.

Alternatively, \eqref{eq:initial_level1} is equivalent to:
\begin{equation}\label{eq:initial_level2}
    \begin{bmatrix}
        -P(h) & e_k \\ e'_k & -\bar{x}_k^2
    \end{bmatrix}< 0~\forall k\in\mathcal{N}
\end{equation}

After using appropriate similarity transformation $\operatorname{diag}(R',1)$ on the left and its transpose on the right, it is sufficient to use \eqref{eq:omega1} to guarantee these conditions.

In order to impose the local conditions for the time-derivative, we consider
\begin{align}
    \dot{h}^2_\upsilon &\leq \phi_\upsilon^2 \notag \\
    \dfrac{\dot{h}^2_\upsilon}{\phi_\upsilon^2} &\leq 1.
\end{align}
With that in mind, by imposing
\begin{equation}\label{eq:bounded_lyap}
     \dfrac{\dot{h}^2_\upsilon}{\phi_\upsilon^2} \leq x'P(h) x,~\forall \upsilon \in\mathcal{R}.
\end{equation}
we ensure that the local conditions for the time-derivative will be valid for $x \in \Omega^*(1)$.

Notice that:

\begin{align}\label{eq:hdot_exploded}
    \dot{h}_\upsilon & = \nabla h_\upsilon \dot{x} \nonumber \\
                    & =\nabla h_\upsilon  [A(h)+B(h)K(h)+B(h)L(\dot{h})]x \nonumber \\
                    &=\nabla h_\upsilon  [A(h)+B(h)K(h)+B(h)\sum^r_{w=1}\dot{h}_w L_w]x \nonumber \\
                    & =\nabla h_\upsilon  [A(h)+B(h)K(h)+B(h)\sum^r_{\substack{w=1\\ w\neq \upsilon}}\dot{h}_w L_w]x + \nabla h_\upsilon \dot{h}_\upsilon B(h)L_\upsilon x \nonumber \\
    \dot{h}_\upsilon       &= (1 -\nabla h_\upsilon B(h)L_\upsilon x )^{-1}\nabla h_\upsilon  [A(h)+B(h)K(h)+B(h)\sum^r_{\substack{w=1\\ w\neq \upsilon}}\dot{h}_w L_w]x
\end{align}
    
Substituting \eqref{eq:hdot_exploded} into  \eqref{eq:bounded_lyap} it follows that:

\begin{equation}\label{eq:bounded_lyap_exploded}
    x'\left(\dfrac{\Pi'(h)\Pi(h)}{\phi_\upsilon^2} - P(h) \right) x <0
\end{equation}
\noindent where:
\begin{equation}
    \Pi(h) = (1 -\nabla h_\upsilon B(h)L_\upsilon x )^{-1}\nabla h_\upsilon  [A(h)+B(h)K(h)+B(h)\sum^r_{\substack{w=1\\ w\neq \upsilon}}\dot{h}_w L_w
\end{equation}

Applying Schur complement to \eqref{eq:bounded_lyap_exploded} results in:
\begin{equation}
    \begin{bmatrix}
    -P(h) & \bullet \\
    A(h)+B(h)K(h)\sum^r_{\substack{w=1\\ w\neq \upsilon}}\dot{h}_w L_w & 
    -\phi_\upsilon^2(1 -\nabla h_\upsilon B(h)L_\upsilon x )^2
    \end{bmatrix}<0
\end{equation}

In order to deal with the term $\nabla h_\upsilon B(h)L_\upsilon x$, we will consider that $\left|1 - \nabla h_\upsilon B(h)L_\upsilon x\right| \geq \mu_\upsilon$, so that $-(1 -\nabla h_\upsilon B(h)L_\upsilon x )^2 \leq -\mu_\upsilon^2$ and a sufficient condition becomes
\begin{equation}
    \begin{bmatrix}
    -P(h) & \bullet \\
    A(h)+B(h)K(h)\sum^r_{\substack{w=1\\ w\neq \upsilon}}\dot{h}_w L_w & 
    -\phi_\upsilon^2\mu_\upsilon^2
    \end{bmatrix}<0,
\end{equation}
which, from a similarity transformation done by multiplying it with $\operatorname{diag}(R',1)$ left and its transpose on the right, can be ensured by \eqref{eq:qii} and \eqref{eq:qij}.

We need thus to enforce that $\left|1 - \nabla h_\upsilon B(h)L_\upsilon x\right| \geq \mu_\upsilon$, which can be checked by $(\nabla h_\upsilon B(h)L_\upsilon x)^2 \leq (1 - \mu_\upsilon)^2$. Similarly to what we did before, we can ensure that this is satisfied for $x \in \Omega^*(1)$ by imposing that
\begin{align}
    \dfrac{x' L_\upsilon' B(h)' \nabla h_\upsilon' \nabla h_\upsilon B(h)L_\upsilon x}{1 - \mu_\upsilon
    ^2} \leq V(x)
\end{align}
which, from a Schur's complement, leads to
\begin{align}
    \begin{bmatrix}
        -P(h) & \bullet \\ \nabla h_\upsilon B(h)L_\upsilon & -(1 - \mu_\upsilon)^2
    \end{bmatrix} \leq 0.
\end{align}
By employing a similarity transformation, done by multiplying it with $\operatorname{diag}(R',1)$ left and its transpose on the right, this last condition can be ensured by \eqref{eq:Xiuq}.

Finally, in order to enlarge $\Omega^*(1)$, we consider that the Lyapunov function is smaller than a quadratic function
\begin{align}
    x' P(h) x \leq x' H^{-1} x \label{eq:lyapQuad}
\end{align}
and maximize the volume of the 1-sublevel set of this quadratic function, which will be inside of $\Omega^*(1)$, by maximizing $\log(\det(H))$. In order to enforce \eqref{eq:lyapQuad}, we have
\begin{align}
    - H^{-1} + P(h) &\leq 0 \\\
    -R' H^{-1} R + T(h) &\leq 0 \\
    -R-R'+H+ T(h) &\leq 0
\end{align}
in which the last inequality comes from the fact that $-R' H^{-1} R \leq -R-R'+H$. By taking the vertices over the summation of $T(h)$, this leads to the conditions in \eqref{eq:lyapQF}.
\end{proof}

\subsubsection{Simulation Example}

Consider a TS fuzzy model \eqref{eq:TS-model} with the parameters presented in \cite[Example 7]{Lee:Inf:2012}:
\[
    A_1 =
    \begin{bmatrix}
    4 & -4 \\ -1 & -2
    \end{bmatrix}, ~~
    A_2 =
    \begin{bmatrix}
    -2 & -4 \\ 20 & -2
    \end{bmatrix},
    B_1 =
    \begin{bmatrix}
    1 \\ 10
    \end{bmatrix}, ~~
    B_2 =
    \begin{bmatrix}
    1 \\ 1
    \end{bmatrix},
\]

The membership functions are $h_1(x_1) = \frac{1 + \sin(x_1)}{2}, h_2(x_1) = \frac{1 - \sin(x_1)}{2}$, with $x_1 \in [-2, 2]$ and $x_2 \in [-1.35\pi, 1.35\pi]$.

Since the gradient of $h_1$ and $h_2$ are given by
\begin{align*}
    \nabla h_1 &= \begin{bmatrix}  0.5 \cos(x_1) & 0\end{bmatrix} \\
    \nabla h_2 &= \begin{bmatrix}  -0.5 \cos(x_1) & 0\end{bmatrix}
\end{align*}
we have that
\begin{align*}
    \zeta_1^1 = \begin{bmatrix}  0.5 & 0\end{bmatrix},~ 
    \zeta_1^2 = \begin{bmatrix}  -0.5 & 0\end{bmatrix},~ 
    \zeta_2^1 = \begin{bmatrix}  -0.5 & 0\end{bmatrix},~ 
    \zeta_2^2 = \begin{bmatrix}  0.5 & 0\end{bmatrix} 
\end{align*}

In this example the objective is to find locally stabilizing PDC controllers with the largest estimated domain of attraction, comparing the conditions from Theorem \ref{theo:local_theo} with and without the derivative term in the control law. Therefore, an heuristic approach was employed to find the best parameters for each case. For the conditions with the derivative term, we considered $\alpha= 0.006$, $\underline{\phi}_1 = \underline{\phi}_2 = 28.5$, and $\mu_1 = \mu_2 = 0.83$. For the conditions without the derivative term, we considered $\alpha= 0.016$, $\underline{\phi}_1 = \underline{\phi}_2 = 12$. Once again, the codes are implemented using the combination of MatLab, YALMIP, and SeDuMi. 

The controller matrices found, for the control law with the derivative term, were:
\begin{align*}
    K_1 &=
    \begin{bmatrix}
    13.755 & -11.2376 
    \end{bmatrix}, ~~
    K_2 =
    \begin{bmatrix}
    14.9228 & -14.5855
    \end{bmatrix}, \\
    L_1 &=
    \begin{bmatrix}
    -0.1496 & 0.1481
    \end{bmatrix}, ~~
    L_2 =
    \begin{bmatrix}
    0.1496 & -0.1481
    \end{bmatrix}.
\end{align*}

The Lyapunov function matrices found, for the control law with the derivative term, were:
\[
    P_1 =
    \begin{bmatrix}
    0.3621 & -0.1559 \\ -0.1559 & 0.1227
    \end{bmatrix}, ~~
    P_2 =
    \begin{bmatrix}
    0.2596 & -0.1146 \\ -0.1136 & 0.1296
    \end{bmatrix}.
\]

\begin{figure}[h!]
\centering
\includegraphics[width=.725\textwidth]{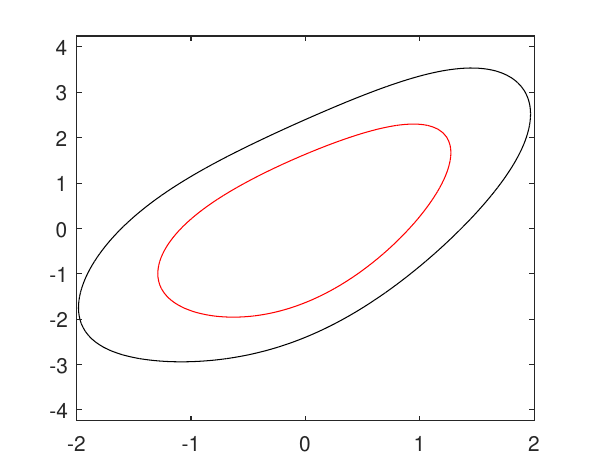}
\caption{Maximum DOA for the local stabilization conditions: red set  is provided by conventional PDC controller, whereas the black set is provided by Theorem~\ref{theo:local_theo}.}
\label{fig:compare_DOA}
\end{figure}

\begin{figure}[h!]
\centering
\includegraphics[width=0.725\textwidth]{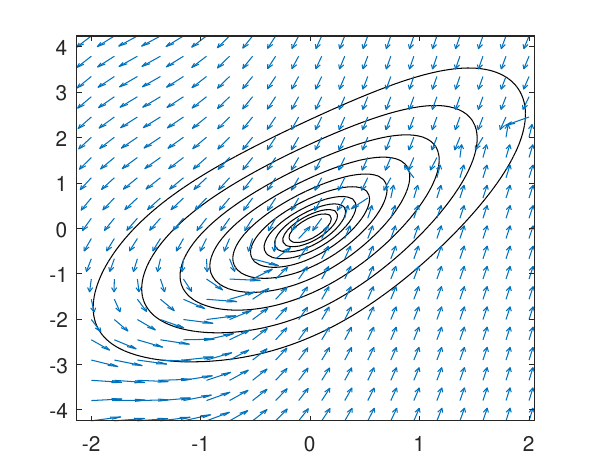}
\caption{Level set curves for the computed Fuzzy Lyapunov Function (black contours). The arrows represent the phase space trajectories, which are always pointing inwards the curves, as preconized by the Lyapunov stability.}
\label{fig:quiver}
\end{figure}

Figure~\ref{fig:compare_DOA} show the maximum estimate of the domain of attraction provided by Theorem~\ref{theo:local_theo} with and without the derivative term in the control law. The analysis reveal that the area provided by the proposed approach is considerably larger, almost doubling the area of the guaranteed domain of attraction. Figure~\ref{fig:quiver} shows the phase portrait of the controlled system, by using blue arrows that indicated $\dot{x}$ for some states. It also shows the level set contours (sets in black color) for the FLF obtained. As expected, since this is indeed a Lyapunov function, the arrow are always pointing inwards the level curves.

\begin{figure}[h!]
\centering
\includegraphics[width=0.725\textwidth]{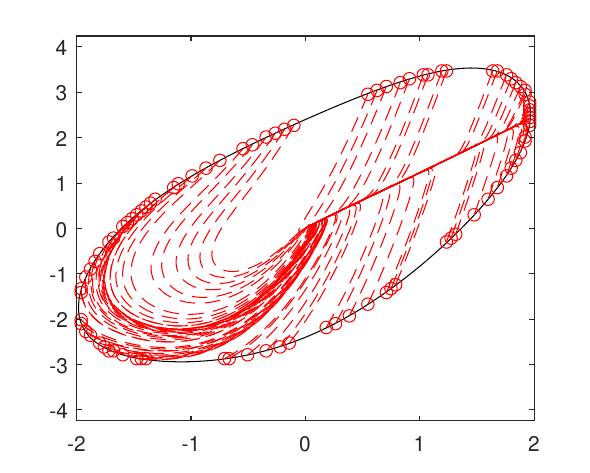}
\caption{Some trajectories are represented by red solid curves and their respective initial conditions are represented by red circles. The black set indicates the best estimate of DOA. Notice that all trajectories starting inside this set remain within the set.}
\label{fig:traj}
\end{figure}

Finally, some borderline trajectories are depicted in Figure~\ref{fig:traj}, i.e., trajectories that start at the limit of the DOA. It is possible to see that all trajectories beginning inside the computed DOA remain inside of it. It is also interesting to witness that the proposed controller enforces distinct behaviors to state convergence towards the origin. In the second and third quadrant, the trajectories are more oscillatory, whereas in the remaining they are more monotonic, converging to a central line that flows directly to the orign.

\section{Conclusions and Future Works}\label{sec:conclu}

Resorting to the recent advancements in FLF literature, we proposed a new type of fuzzy controller that uses explicitly the information of the time-derivative of the membership grades in the control law. This new topology includes the PDC controller as special case and numerical examples have shown that its superiority over recent results.

As future works, we intend to explore the real-time implementation of this strategy and also consider its application to output feedback, observer/estimator design and to include performance criteria.

\bibliographystyle{abbrvnat}
\bibliography{refs}

\end{document}